# Surface Engineering of Synthetic Nanopores by Atomic Layer Deposition and Their Applications

Ceming Wang[1], Delin Kong[2], Qiang Chen[2*], Jianming Xue[1*]

1- State Key Laboratory of Nuclear Physics and Technology, School of Physics, Peking University, Beijing 100871, P. R. China;

2-Laboratory of Plasma Physics and Materials, Beijing Institute of Graphic Communication, Beijing 102600, P. R. China

**ABSTRACT**：In the past decade, nanopores have been developed extensively for various potential applications, and their performance greatly depends on the surface properties of the nanopores. Atomic layer deposition (ALD) is a new technology for depositing thin films, which has been rapidly developed from a niche technology to an established method. ALD films can cover the surface in confined regions even in nano scale conformally, thus it is proved to be a powerful tool to modify the surface of the synthetic nanopores and also to fabricate complex nanopores. This review gives a brief introduction on nanopore synthesis and ALD fundamental knowledge, then focuses on the various aspects of synthetic nanopores processing by ALD and their applications, including single-molecule sensing, nanofluidic devices, nanostructure fabrication and other applications.

## 1. Introduction

Nanopores have attracted a great deal of scientific interests due to their prospective application in sensing biomolecules. In particular, nanopore-based sensors may offer new opportunities for a fast and low-cost DNA sequencing method that can be scaled for high-throughput DNA analysis. Recent achievements suggest that nanopore-based sensors could be competitive with other third-generation DNA sequencing technologies, and may be able to rapidly and reliably sequence the human genome for under $1,000[1].

Currently, most experiments on nanopores have concentrated on biological nanopore α-haemolysin[2] and on artificial pores in solid-state membranes[3]. Although biological pores have been proved to be very useful for a range of interesting translocation experiments, they exhibit a number of disadvantages such as limited lifetime, low dynamic voltage range (beyond which the electrical breakdown of the supporting lipid bilayer will occur) and lack of mechanical stability. Fabrication of nanopores from solid-state materials presents obvious advantages over their biological counterpart such as very high stability, controllable of diameter and channel length, and the potential for integration into devices and arrays. Though solid-state nanopores show much promise, major hurdles still remain. The methods to create nanopores usually yield uncharacterized and unfavorable surface properties interfering with the pore sensing abilities[4,5]. Thus, it is crucial to improve the surface properties of synthetic nanopores for a better performance in sensing biological molecules.

Solid-state nanopores are also widely used as an important class of nanostructures in nanofluidics, which has experienced considerable growth in recent years with significant scientific and practical achievements[6]. For the successful development and investigation of nanofluidics, reliable fabrications of nanopores with special functionalities are essential. Examples include fabrication of electrode-embedded nanopores to efficiently manipulate the ion transport in nanopore[7], and the control of surface charge to build nanofluidic circuit elements, such as diodes and transistors[8].

Another class of synthetic nanopores can be found in multi-pore membranes. Anodic aluminum oxide (AAO) and track-etched membrane are of major interests because they are relatively cheap to manufacture and are well suitable for up scaling. These materials are widely used as templates for the synthesis of nanowires and nanotubes[9]. Recent advances have shown that very promising functional nanoporous material can potentially be achieved by modifying the nanoporous materials with surface engineering technology.

Atomic layer deposition (ALD), as a powerful surface engineering tool, has attracted the attention of researchers in the field of nanopore. The outstanding features of ALD, such as precise thickness control and highly conformal deposition, allow for new strategies in the modification of chemical and physical properties of nanopores and synthesis routes to novel nanostructures.

ALD is a self-limited film growth technique, which is characterized by alternating exposure of the growing film to chemical precursors, resulting in the sequential deposition of monolayers[10]. ALD was invented in the 1970s, and further developed in the 1980s for depositing insulator films such as ZnS and $Al_2O_3$ in purpose of electroluminescent flat panel displays. Nowadays ALD finds its new application in microelectronics industry, e.g. for growing high-*k* gate oxides. It can deposit many kinds of materials, including oxides, nitrides, sulfides as well as metals[11].

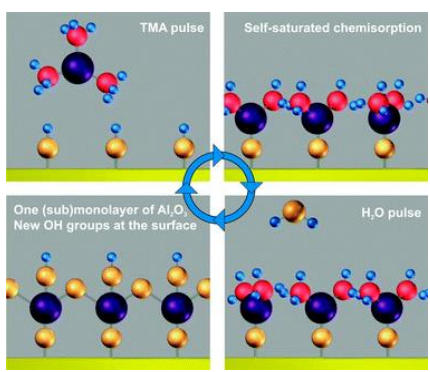

**Fig. 1** Principle of ALD, illustrated by the process for deposition of $Al_2O_3$ using TMA and $H_2O$. Small blue spheres represent hydrogen atom, red spheres for carbon, large violet spheres are for aluminum, and yellow spheres for oxygen atoms. Figure courtesy of Cambridge NanoTech Inc.

As an example, we discuss the growth of a $Al_2O_3$ film using ALD[11]. The basic ALD cycle for $Al_2O_3$ deposition is illustrated in Fig. 1. In the beginning, as shown in the top left, the substrate terminated by hydroxyl groups is exposed to a pulse of trimethyl aluminum vapor (TMA). The TMA reacts with the hydroxyl groups on all exposed surfaces in the sample until all accessible hydroxyl groups have been consumed. TMA does not react with itself so that only a single layer of TMA can be adsorbed (Fig. 1, top right). The TMA dose is followed by an evacuation of the reaction chamber through purging and pumping, then a pulse of e.g. water vapor, or ozone is introduced (Fig. 1, bottom right) into the chamber. The water vapor reacts with the adsorbed TMA, and hydrolyses the residual methyl groups. This surface reaction results in the formation of a monolayer alumina. At last, as shown in the bottom left of Fig. 1, hydroxyl groups are presented to terminate the first alumina layer. Therefore, the ALD process can run repeatedly to deposit atomic layer film one by one.

One of the obvious advantages of ALD is it can precisely control the film thickness at monolayer level. Besides, the self-limiting aspect of ALD leads to excellent step coverage and conformal deposition on high aspect ratio structures. The conformal deposition is very important because many of the nanopores that will be described are characterized by high-aspect-ratio. There have been several studies in the literature which attempted experimentally and/or theoretically to understand the film deposition characteristics in nanopores or trenches[12-14]. A simple theory proposed by Gordon et al.[13] could generate conditions for which full step-coverage could be attained within narrow high-aspect-ratio holes. This theory depends on the exposure time and partial pressure of the precursor species, which limits conformality (deep penetration into the nanopore) in the ALD process. Larger pressures or exposure times will allow the reactant to penetrate deeper into the nanopores, whereas a larger molecular mass will hinder its ability to enter a nanopore.

This paper will give an overview of the recently evolved works on the applications of ALD in surface engineering of synthetic nanopore. The first part of the article presents a short introduction to the fabrication of nanopores. Then the focus is on the various aspects of synthetic nanopores processing by ALD and their applications, including single-molecule sensing, nanofluidic device, nanostructure fabrication and other applications.

## 2. Synthetic nanopore fabrications

Synthetic nanopores are being fabricated using various methods[15]. The past decade has seen major development of single synthetic nanopores, aiming to address the limitations of biological pores. Several fabrication methods of single synthetic nanopores as well as nanoporous materials have been demonstrated, and these are described below. In general, each fabrication method is tied to a particular substrate material.

### 2.1 Ion beam sculpting

The first sythetic nanopore used for DNA sensing was developed by Golovchenko and co-workers at Harvard university[16]. They reported a novel technique, ion beam sculpting, by which they fabricated single nanopores in thin SiN membranes with a true nanometer control. They prepared a flat $Si_3N_4$ surface containing a bowl-shaped cavity on one side, and impinged on the opposite surface with argon ion beams (Fig. 2, left). As material is removed from the flat $Si_3N_4$ surface, the surface will ultimately intercept the bottom of the cavity, which causes the pore opened. By implementing a feedback-controlled ion sputtering system, they could extinguish the erosion process upon breakthrough and thus create a molecular-scale pore as small as 1.8 nm.

Although a nanopore can be created from a cavity in the membrane under conditions where the sputtering erosion process dominates, the ion beam can also be used to stimulate the lateral transport of matter, which causes a pre-existing larger pore, such as a focused ion beam (FIB) drilled pore, to closed (Fig. 2, right)[17]. By changing the sample temperature or ion beam parameters during the process, it is possible to control pores opening or closing dominantly, obtaining the fine-tuning of pores in nanometer range.

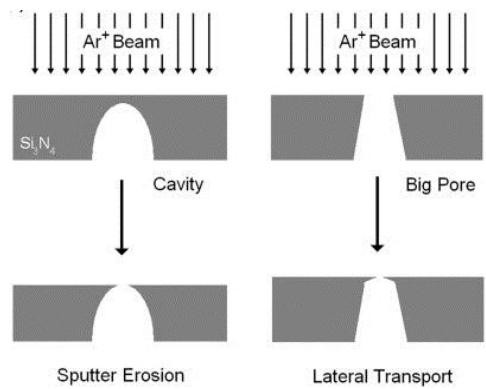

**Fig. 2** Ion bean sculpting to make nanopores from a cavity (left) or from a through hole (right). Either sputter erosion or lateral transport processes dominate, depending on the selected conditions used in the ion beam sculpting apparatus[16], Reproduced with permission from Ref.[15]

**2.2 Electron beam sculpting**

Stormet et al. pioneered the use of high-energy electron beams to fine-tune the size of silicon oxide nanopores[18], it then became one of the most popular methodologies to fabricate small nanopores. In their approach (Fig. 3), free-standing membranes of Si, SiN, or $SiO_2$ can be made based on by Si microfabrication techniques. Subsequently, holes can be fabricated through using electron-beam lithography and subsequent etching. A typical diameter of such holes is around 20 nm. In imaging these holes with a transmission electron microscope (TEM), they discovered that high-intensity wide-field illumination with electrons slowly modified the nanopore size. Large holes, with a diameter greater than the membrane thickness, grew in size, whereas small holes were shrunk. Note that this provides a way to fine-tune the nanopores to a small size with subnanometer resolution and with direct visual feedback from a commercial TEM. This process of glassmaking at the nanoscale occurs as the $SiO_2$ membrane flows to decrease the surface tension and so slowly reshapes around the hole[19]. If electron-beam exposure is stopped, the structure will be entirely frozen and stable. It has also been shown that a pore can be directly drilled in a membrane by a locally focused electron beam in a TEM[18, 20], making laborious preparatory electron-beam lithography of larger nanopores unnecessary. Again, such pores can be modified with the wide-field TEM illumination.

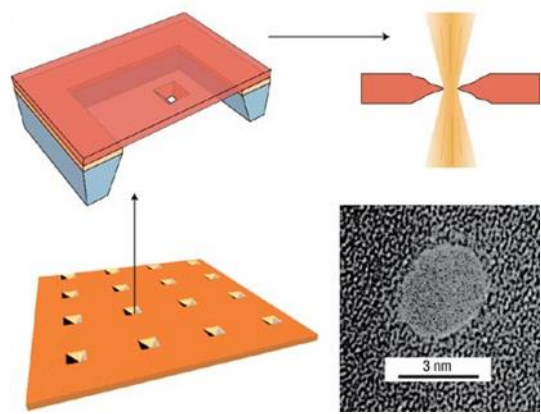

**Fig. 3** The TEM technique developed by Stormet et al.[18]. An electron beam drilled a hole in the membrane. The result can be directly monitored in the electron microscope and the electron beam can also be used to enlarge or shrink the nanopore in a controlled way. The figure at the bottom right shows a typical TEM image of a nanopore (with a diameter of 3 nm in this example). Reproduced with permission from Ref.[3]

**2.3 Ion track etching**

The track-etching method[21] has been used to produce single nanopores in polymer membranes. The single-pore fabrication has similar roots to that for producing track-etched membranes. These contain large numbers of nanoscale pores, prepared by irradiating membranes with a beam of energetic heavy ions and then chemically etching the resultant ion tracks, each ion creates a single track that can be preferentially etched. The principle of track-etching technique is shown in Fig. 4. A modified irradiation process developed at the GSI (Darmstadt, Germany) allows the fabrication of single-pore membrane. This is achieved by tuning the intensity of heavy-ion irradiationto limit the number of passing ions to one[22].

The size and shape of the pores are controlled in the etching step by varying the type and concentration of the etchant, the reaction temperature, and the duration of etching process. Conically shaped single nanopores have attracted special interest. Compared to a cylindrical pore of the same limiting diameter, a conical pore has a much lower electrical resistance and is therefore easier to analyze. The method of obtaining conically shaped nanopores is based on asymmetric etching of irradiated films[23]. In the case of poly(ethylene terephthalate) (PET) and polycarbonate (PC) membranes, conically shaped pores are obtained by placing an irradiated foil between a concentrated base on one side and an acidic solution on the other[23].

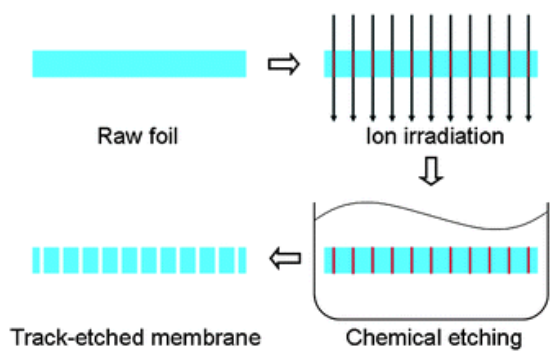

**Fig. 4** Schematic of the fabrication process of track-etched membrane by the track-etching technique.

**2.4 Nanoporous materials**

Numerous methods have been developed to fabricate nanoporous films and materials[24]. Here we limit the description to materials that have been already modified with ALD. The track-etching method is one which has been used to produce commercially viable nanopore arrays with diameters as low as ~10 nm and packing densities as high as $6\times10^8$ pores /cm$^2$. Nanopore arrays can also be fabricated through an anodization process of thin aluminum films. Anodized aluminum oxide (AAO) has been one of the substrates of choice for the fabrication of nanoporous materials. AAO membranes are fabricated by anodic oxidation of an aluminum substrate in acidic solutions. The resulting membrane consists of densely packed hexagonal pores of 10–200 nm in diameter, whose size depends on the oxidation conditions. Architectures consisting of long-range ordered nanochannel arrays can be fabricated out of AAO on the millimeter scale[25].

## 3  ALD into solid-state nanopores
### 3.1 Single-molecule sensing

For the past decade, solid-state nanopores have been developed as a powerful technique for sensing biological macromolecules[3]. It is evident that both the membrane surface properties and the nanopore dimension are critical. For example, the surface properties of the pore and its immediate surroundings should not repel the molecules detected; the aperture of the pores must have a diameter large enough to allow the molecules to translocate, also should be small enough to optimize signal response to the molecule presence. There are impediments to achieving simultaneous control of both surface properties and nanoporesize because the choice of membrane material is usually limited by the technical features of the fabrication processes.

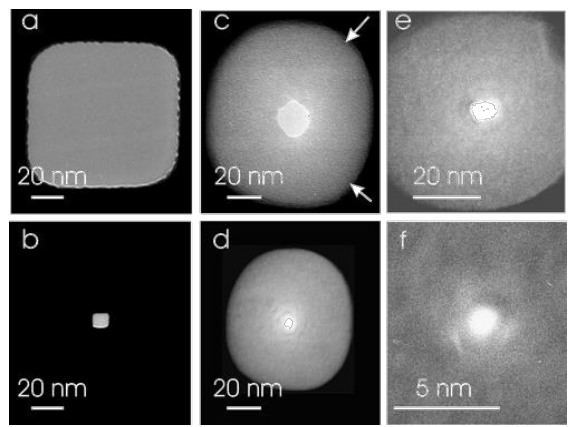

**Fig.5** Transmission electron microscopy images of several pores before (top row) and after (bottom row) deposition of Al$_2$O$_3$ coatings by atomic layer deposition. (Left) 500 (Center) ,70 (Right) , and 24 cycless of Al$_2$O$_3$, respectively. Reproduced with permission from Ref.[4].

Chen et al. firstly suggested that ALD can afford a finishing step, to finely tune the sizes and surface properties of nanopores fabricated by ion beam sculpting[4]. ALD was proved to be highly conformal providing a uniform coating to all exposed surfaces including the surface inside nanopores, therefore the pore maintained its initial shape and only reduced its size, as shown in Fig. 5. By using ALD, the diameter of a nanopore can be fashioned with atomic precision, shrinking an oversized pore to a preferred smaller diameter. In particular, nanopores with various diameters were easily produced, based upon the calibrated deposition rate of $0.099 \pm 0.012$ nm per reaction cycle.

ALD can also adjust the surface properties by choosing suitable coating material. Deposition of Al$_2$O$_3$ can enhance the molecule sensing characteristics of fabricated nanopores by fine-tuning their surface properties[4]. For example, sculpting produces pores with varing amounts of negative surface charge, and the frequency of DNA translocations through pores with high surface charge was very low, thought to be due to electrostatic repulsion and opposing electroosmotic flow. Al$_2$O$_3$ coating eliminated the pore surface charge that repel molecules to be detected, restored a high frequency of DNA translocations and also reduced the level of 1/f noise. Moreover, Al$_2$O$_3$ is a thermally and chemically stable, insulating, dielectric material that inhibits direct electron tunneling.

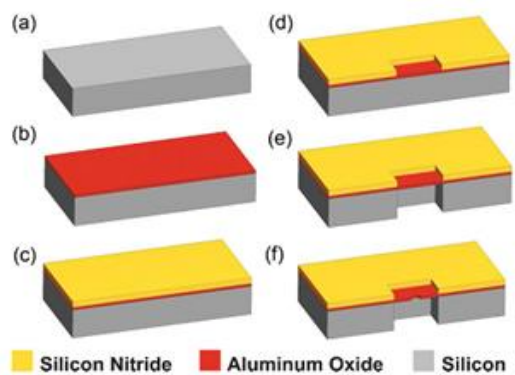

**Fig. 6** Process flow for the formation of $Al_2O_3$ nanopores. a) start with double-side polished 300 mm thick silicon wafer. b) deposit 70 nm of $Al_2O_3$ by ALD. c) deposit 500 nm low-stress SiN using PECVD process. d) pattern 30 mm×30 mm windows on the wafer front side via optical lithography and DRIE. e) pattern 30 mm×30 mm windows on the wafer backside and etch using DRIE and stop on the $Al_2O_3$ layer creating a membrane. f) use a tightly focused electron beam to form nanometer-sized pores. Reproduced with permission from Ref.[5]

Bashir's group reported a different approach via ALD to fabricate a $Al_2O_3$ nanopore for the detection of single DNA molecules[5]. The process flow for the fabrication of $Al_2O_3$ nanopores is outlined in Fig. 6. ALD was used to deposit 70 nm of $Al_2O_3$. The self-limiting growth characteristic of ALD enables excellent uniformity over large areas, accurate control of film composition and thickness. Plasma-enhanced chemical vapor deposition (PECVD) was used next to deposit 500 nm of low stress SiN as a passivation layer to help reducing device capacitance and electrical noise. They used a tightly focused electron beam from a field-emission-gun transmission electron microscope to form nanopores through electron beam induced decompositional sputtering. These $Al_2O_3$ nanopores exhibit enhanced mechanical properties (increased hardness, decreased stress) and improved electrical performance (low noise, high signal-to-noise ratio) over previously reported $SiO_2$ and $Si_3N_4$ nanopores. For example, the high frequency (f >10 kHz) noise performance of $Al_2O_3$ nanopores shows an order of magnitude lower than the existing $Si_3N_4$ and $Al_2O_3$-coated $Si_3N_4$ structures [4, 26], resulting in high sensitivity and exceptional signal-to-noise performance.

Interestingly, they also observed drastic changes in the material properties of the $Al_2O_3$ nanopore which significantly impacting the sensitivity of these nanoscale single-molecule sensors[27]. During electron-beam irradiation, the local nanostructure and morphology of the $Al_2O_3$ pore transform from an amorphous, stoichiometric structure (O to Al ratio of 1.5) to a heterophase crystalline network, deficient in O (O to Al ratio of 0.6). Direct metallization of the pore region is observed during irradiation, thereby permitting the potential fabrication of nanoscale metallic contacts in the pore region with application to nanopore-based DNA sequencing. Dose-dependent phase transformations to purely γ and/or α-phase nanocrystallites are also observed during pore formation, allowing for surface charge engineering at the nanopore/fluid interface. Fast translocation velocity of DNA has been identified as one of the critical issues in DNA sequencing. The translocation of dsDNA through these nanometer-sized alumina pores revealed average translocation velocities that were an order of magnitude smaller than that observed in $Si_3N_4$ and $SiO_2$ systems under similar conditions, attributed to strong DNA–nanopore interactions. These intereaction influenced DNA translocation kinetics include electrostatic binding of anionic DNA to the positively charged nanopore surface enhanced by γ and α- $Al_2O_3$ nanocrystallite formation and hydrophobic polymer–pore interactions promoted by the relatively high surface roughness of electron-beam-irradiated $Al_2O_3$. High DNA translocation velocities (~30 bases/ms)[4] limit the utility of conventional $SiO_2$ and $Si_3N_4$-based nanopore technologies in high-end DNA sensing and analysis applications, including single-nucleotide detection.

This fabrication process proposed by Bashir et al[5,27] can potentially allow for the formation of ultrathin membranes (thickness <10 nm) due to the precise thickness control offered by ALD. This is particularly useful in forming a solid-state analog to the lipid bilayer (thickness ~ 4–5 nm)[28], an important tool in better understanding the kinetics governing biomolecule transport through proteinaceous pores in cellular membranes.

Currently, most sythethic nanopore for single molecule sensing are straight which aligned normal to the membrane surface. Chen et al. in University of New Mexico reported a 'kinked' silica nanopores for DNA sensing[29]. The orientations of these 'kinded' nanopores deviate periodically from the surface normal. They find that these kinked nanopores reduce the translocation velocity up to five-fold compared with comparably sized straight-through pores. This nanopore array can exhibit nearly perfect selectivity for ssDNA over dsDNA. In their experiment, they used ALD to adjust the pore diameter from 2.6 to 1.4 nm (which are between the diameters of dsDNA and ssDNA) as well as altering the surface chemistry. Pore-size reduction by ALD occurs uniformly for all pores, allowing selectivity to be developed in an array format. Their results, showing nearly perfect selectivity for ssDNA over dsDNA in ALD-modified nanopores, could expand the range of potential applications of synthetic solid-state nanopores in biotechnology by tuning the pore size to be between the diameters of dsDNA and ssDNA.

### 3.2 Nanofluidic device

The transport of fluid in and around nanometer-sized objects with at least one characteristic dimension below 100 nm enables the occurrence of phenomena that are impossible appearance at bigger length scales. The proximity of this dimension and the Debye length, the size of biomolecules such as DNA or proteins, or even the slip length, added to the excellent control on the geometry gives unique features to nanofluidic devices[6]. This research field was only recently termed nanofluidics, but it has deep roots in science and technology. Nanofluidics has experienced considerable growth in recent years, as is confirmed by significant scientific and practical achievements[6].

Electrical manipulation of charged species such as electrolyte ions, DNA, proteins, and nanoparticles is an important task in nanofluidic systems. In this regard, the combination of "Electronics" and "Nanofluidics" leads to the field of "Electrofluidics", which utilizes the electrical behaviors of fluids for solid state device applications. The deposition of a dual layer, i.e. metal–oxide, by ALD deposition of a dielectric layer over the metallized nanopores, allows the metal to be isolated from its surrounding environment, such as an electrolyte, to form a metal–insulator–electrolyte structure, also called a nanofluidic transistor[8].

Nam et al. reported a method to fabricate electrode-embedded multiple nanopore structures with a sub-10 nm diameter, which was designed for electrofluidics applications such as ionic field effect transistors[30]. They used a membrane consisting of a conducting layer (TiN 30nm) sandwiched by dielectric films ($Si_3N_4$ 20 nm). Then, a nanopore in this membrane was generated by electron beam lithography. To obtain sub-10 nm nanopores, they coated the intial nanopore with $TiO_2$ by self limiting ALD process (Fig. 7). They successfully demonstrated that 70∼80 nm of diameter pores can be shrunk down to sub-10 nm diameter. In this device, the nanopore ionic channels surrounded by gate dielectric ($TiO_2$)/gate electrode(TiN) allows three-dimensional gating, which showed highly efficient controllability on channel conductivity than other structures. Moreover, the demonstration of gating property of nanopores has a great potential for manipulation of ions and biomolecules in integrated nanofluidic devices.

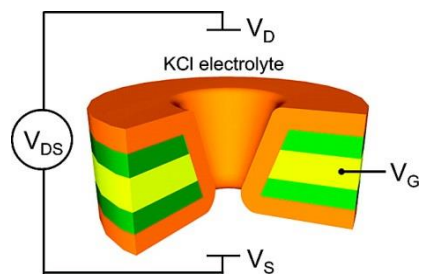

**Fig. 7** Principle of electrode-embedded nanopores. Ionic transport through the nanopores can be manipulated by the surrounding gate dielectric ($TiO_2$, orange) which were coated by ALD and gate electrode (TiN, yellow) (The green color shows $Si_3N_4$). Reproduced with permission from Ref.[30].

These sub-10 nm nanopore structures can also be obtained in a novel self-sealing way as shown in another work by Nam et al.[31]. Their fabrication methods depended strongly on the observed high degree of conformity of ALD films, as well as the precise control of film thickness. Fig. 8 shows schematic figures of the nanochannel fabrication. The fabrication strategy firstly obtained a trench structure in the $Si/SiO_2$ layer (typically 40-60 nm wide). Wet chemical etching of the underlying $SiO_2$ layer through the Si line leads to an undercut geometry, resulting in a roughly circular cross section for the trench. Conformal deposition with ALD uniformly coats the inside of the circular trench (or pipe structure). Continued ALD deposition can result in a pinch-off structure at the top of the pipe, resulting in a buried, open tube. In these fabrication, the ALD process has three advantages: firstly, the conformal film deposition can allow for a self-sealing process of nanochannels during the ALD, if the film thickness is adequate to cover the whole trench region; secondly, the diameter of the nanochannel is automatically determined during the ALD regardless of the deposited ALD film thickness; thirdly, using highly conformal ALD film structures, including $TiO_2$/TiN, and $Al_2O_3$/Ru, nanochannels surrounded by core/shell (high- k dielectric/metal) layers can be fabricated, which are an important functional element in an electrofluidic -based circuit system.

In the context of nanoporous materials, ALD has been widely used as a powerful tool to decorating the pore surfaces of porous membranes[32]. AAO membrane, especially, represents an interesting substrate for the fabrication of functional nanoporous material. Recently, Pardon et al[33] reported the functionalization of nanoporous membranes using ALD. ALD was used to conformally deposit platinum (Pt) and aluminum oxide ($Al_2O_3$) on Pt in nanopores to form a metal–insulator stack inside the nanopore, as shown in Fig. 8. Deposition of these materials inside nanopores allows the addition of extra functionalities to AAO membranes. Conformal deposition of

Pt on such materials enables increased performances for electrochemical sensing applications or fuel cell electrodes. An additional conformal $Al_2O_3$ layer on such a Pt film forms a metal–insulator–electrolyte system, enabling field effect control of the nanofluidic properties of the membrane. This pioneers possibilities for the study of nanofluidic properties as well as for the potential development of nanofluidic biosensors or novel energy harvesting applications using nanoporous devices.

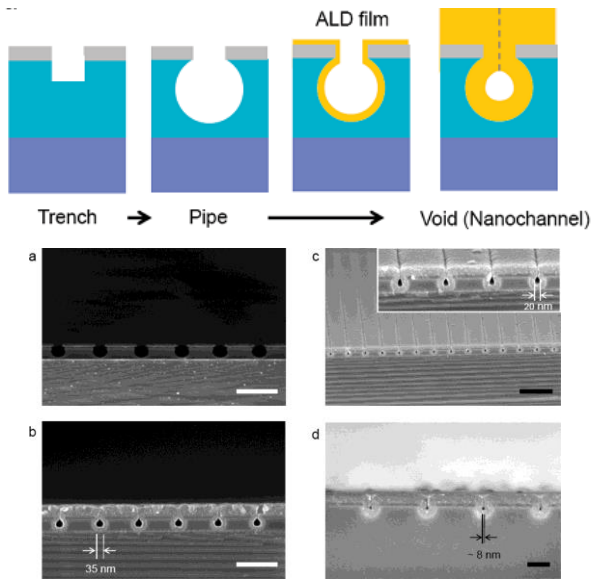

**Fig. 8** Top: The procedure of nanochannel fabrication (gray, amorphous silicon; cyan, $SiO_2$; blue, silicon substrate; yellow, ALD film). Bottom: cross-sectional scanning electron microscope (SEM) images of (a) pipe structures, (b) 35 nm void structures (10 nm wet etching) (c) 20 nm void structures (5 min wet etching), and (d) 8nm void structures (3 min wet etching). (scale bars are (a, b) 250 nm, (c) 500 nm, and (d) 100 nm, respectively), Reproduced with permission from Ref.[31].

### 3.3 Nanostructure fabrication

To a significant extent, the promise of nanotechnology may well be realized in devices and nanostructures that employ: 1) pattern definition through nanofabrication (typically, increasingly challenging lithography); 2) self-alignment, in which conventional processes of material addition or subtraction can be used to form more complex three-dimensional (3D) structures; and 3) nature's tendency toward self-assembly. For self-alingnment approach, the precise control of the thickness of the ALD layers combined to the high step coverage permitted to introduce new approaches for nanostructure fabrication[34].

Porous materials, particularly with one-dimensional high aspect ratio pores, are one of major interests as templates for the synthesis of nanowires and nanotubes[9]. Dense, aligned and uniform arrays of $TiO_2$ nanotubes were fabricated by ALD with the $TiCl_4/H_2O$ process on nanoporous $Al_2O_3$ templates[35] as shown in Fig. 9. The alumina template was then chemically removed to reveal nanotubes with perfectly controllable tube diameter, spacing and wall thickness. In a similar manner ZnO-nanowires were produced by ALD deposition onto a porous alumina template[36] or the pore sizes were reduced by ALD in order to control the thickness of nanowires produced by successive electrochemical deposition[37].

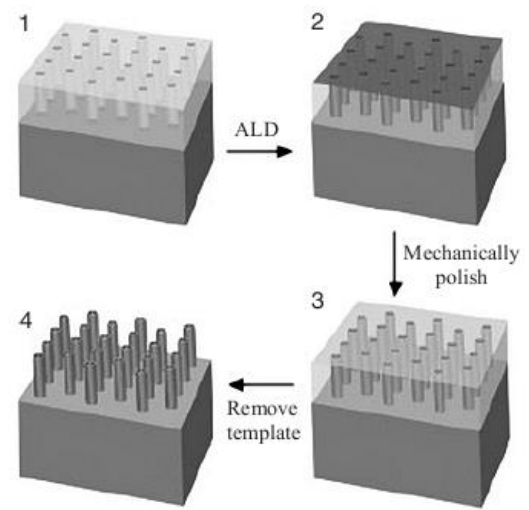

**Fig. 9** Schematic of the process to create titania nanotube arrays on Si substrates. 1) nanoporous-alumina template on a substrate created by anodization of Al film. 2) $TiO_2$ deposited onto the surface of the template by ALD. 3) top layer of $TiO_2$ on alumina removed by gentle mechanical polish. 4) alumina template chemically etched away to reveal array of titania nanotubes on the substrate. Reproduced with permission from Ref.[35].

Other oxide materials fabricated as nanotubes by ALD include $TiO_2$ and $ZrO_2$[38]. In these works a commercially available polycarbonate (PC) track-etched membrane was used as template (AR = 60:1). After chemically etched away of PC, $TiO_2$ and $ZrO_2$ nanotubes could be obtained. The free-standing nanotubes with ca. 30-200 nm diameters could be fabricated in a one-step process because microcontact printed OTS-SAMs prevented deposition onto both sides of the PC template.

Recently, Vandersarl et al. demonstrated that nanotube-based nanomaterial platforms can be used to improve intracellular delivery[39]. In their experiments, $Al_2O_3$ nanotubes/nanostraws were fabricated in a similar manner described above. Detailed fabrication process is illustrated in Fig. 10. The deposited alumina creates the nanostraw bodies within the nanopore interiors and defines the nanostraw wall thickness. Since these nanoscale inter cellular junctions provide fluidic access and promote molecular exchange, which yet are small enough to avoid cell toxicity, the system can be easily extended to deliver many biomolecules, such as membrane-impermeable dyes, plasmids, peptides, RNA, and proteins.

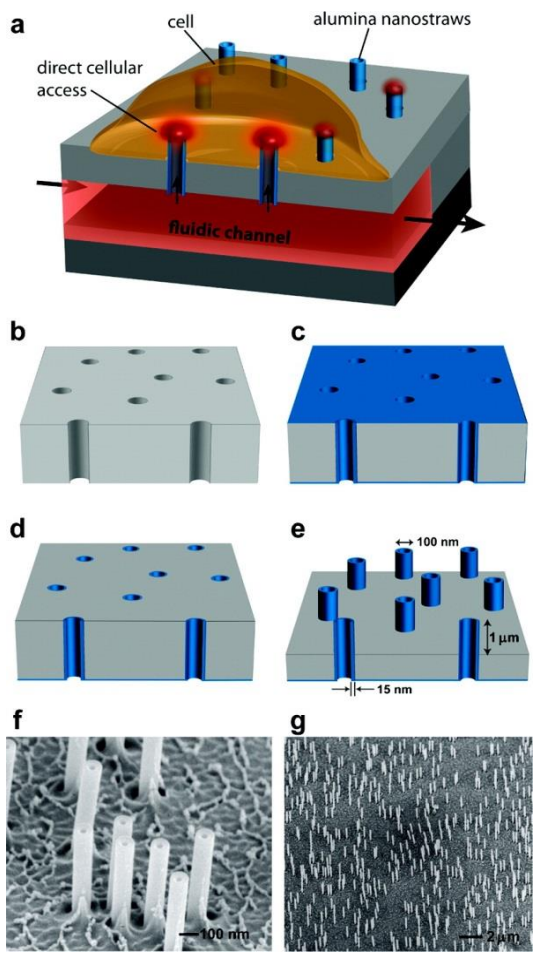

**Fig. 10** Nanostraw−cell interfacing strategy and fabrication. (a) schematic of cell cultured on nanostraw membrane with microfluidic channel access. (b−e) straw fabrication process flow beginning with ananoporous polycarbonate membrane (b), proceeds with conformal alumina atomic layer deposition (c), then an alumina specific directional reactive ion etch (d), and conclusions with a polycarbonate specific directional reactive ion etch (e). (f, g) scanning electron micrographs of nanostraw membranes, Reproduced with permission from Ref.[39].

Nanostructured devices have the potential to serve as the basis for next-generation energy systems that make use of densely packed interfaces and thin films. One approach to making such devices is to build multilayer structures of large area inside the open volume of a nanostructured template. Banerjee et al[40] reported the use of ALD to fabricate arrays of metal–insulator–metal nanocapacitors in AAO nanopores. In their experiment, ALD was used to coat AAO templates with TiN, $Al_2O_3$ and TiN employed as bottom electrode, dielectric and top electrode, respectively, for the fabrication of arrays of metal–insulator–metal nanocapacitors (Fig.11). These highly regular arrays have a capacitance per unit planar area of $\sim 10 \mu Fcm^{-2}$ for 1-mm-thick anodic aluminium oxide and $\sim 100 \mu Fcm^{-2}$ for 10-mm-thick anodic aluminium oxide. This approach are promising to make viable energy storage systems that provide both high energy density and high power density.

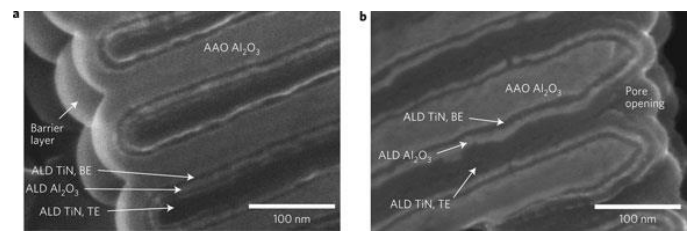

**Fig. 11** a-bottom of tube showing the AAO barrier layer and three layers corresponding to the TiN bottom electrode (BE), $Al_2O_3$ and the TiN top electrode (TE); b-pore openings at the top also show a similar trilayer structure. Thickness measurements at the top and bottom of the pores indicate a step coverage of >93% for all layers. Reproduced with permission from Ref.[40].

### 3.4 Other applications

Nanopores in AAO membrane can be also used as a nanoscale template to test the ALD coating both in experiment and theory[12]. The authors studied the deposition films of $Al_2O_3$ and ZnO in the high-aspect-ratio pores of AAO membranes with ALD mehod. The membranes had pores with diameters of 65 nm, depths spanning the full thickness of 50 mm, and open ends on either side of the membrane. Plain-view of SEM images from the bottom, top, and middle of the $Al_2O_3$ templates deposited by ALD showed a dependence of film conformality on exposure time (Fig. 12). The film thickness recorded from the middle of the template at some depth within a pore varied with exposure time when compared to the thickness recorded near the top and bottom of the templates. The ZnO deposited by ALD in AAO membranes showed similar results. Zinc elemental maps of the cross-sectional profiles of the membranes created by an electron microprobe revealed smaller amounts of Zn near the center of the AAO cross section for shorter exposure times. A simple Monte Carlo simulation was also presented to explain variations with the step-coverage due to insufficient exposure times in both Al$_2$O$_3$ and ZnO cases. The model assumed that molecular transport within the pores was governed by molecular flowing model and that the coatings reacted in a self-limiting manner characteristic of ALD.

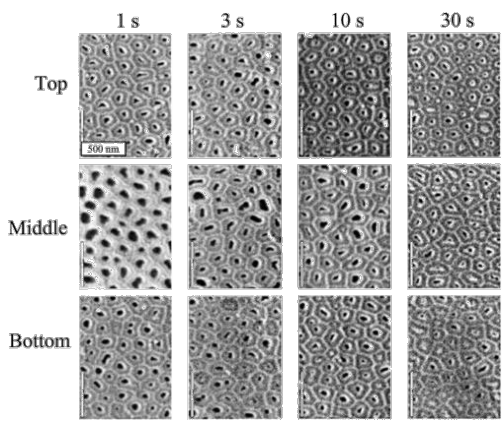

**Fig. 12** SEM images recorded from top, middle, and bottom portions of AAO membranes following 100 cycles of Al$_2$O$_3$ ALD using exposure times of 1, 3, 10, and 30 s, respectively. Reproduced with permission from Ref.[12].

## 4 Conclusions and Outlooks

ALD as a powerful surface engineering method has attracted the attention of researchers in the field of nanopores. The outstanding features of ALD allow for new strategies in the modification of chemical and physical properties of nanopores as well as synthesis routes to novel nanostructures. ALD is a gas phase depostion method based on sequential, self-limiting surface reactions. ALD can deposit very conformal and ultrathin films on high aspect ratio substrates with atomic level control of layer thickness and composition. These unique advantages render ALD an enabling technology for surface engineering of synthetic nanopores. This review has presented a brief introduction to nanopore and ALD. Subsequently, various fabrication methods especially for single solid-state nanopores was introduced. The overview then described various aspects of synthetic nanopores processing by ALD and their applications, including single-molecule sensing, nanofluidic device, nanostructure fabrication and other applications.

The future prospects for surface engineering of sythetic nanopores by ALD are very promising. Various materials can be deposited using ALD techniques. The developments of low-temperature deposition processes (<100 ℃) have significantly enlarged the application range of ALD and the method was already applied on temperature-sensitive nanopore materials, like track-etched polymer nanopores and atomic sensors. The availability of many commercial ALD reactors continues to make ALD accessible for researchers in the nanopore community. The future should see ALD continue to expand into new areas of nanopore researchs and find additional applications that benefit from its precise thickness control and conformality.

For example in single molecule sensing applications, fabricating arrays of uniform solid-state nanopores with diameters less than 10 nm remains a daunting task. The current research-scale fabrication methods are so tedious, slow and manpower-expensive. Thickness control at the Ångstrom provided by ALD can potentially allow for the formation of large-scale sub-10 nm nanopore arrays conbined with other pore formation methods.

In addition, researchers have started using a similar platform for DNA sensing to investigate other analyte molecules such as proteins. Proteins are incredibly diverse and complex in the terms of size, shape, and founction. Particularly, understanding the intereactions between inorganic materials and proteins using nanopore technology is of key importance in various aspects, such as biosensors, biomaterials, and biomedical molecular diagnosis. The various depositing materials provided by ALD can make sythetic nanopore to be a powerful tool for protein-inorganic materials intereaction researchs.

**Acknowledgements**
This work was financially supported by NSFC (No. 11175024).